\documentclass[twocolumn]{aastex62}

\received{2020 April 18}
\revised{2020 July 6}
\accepted{2020 August 7}
\submitjournal{AJ}

\shorttitle{Gaia19bey Outburst}
\shortauthors{Hodapp et al.}

\begin{document}
\turnoffedit
\title{The Outburst of the Young Star Gaia19bey}

\correspondingauthor{Klaus Hodapp}
\email{hodapp@ifa.hawaii.edu}

\author[0000-0003-0786-2140]{Klaus W. Hodapp}
\affil{University of Hawaii, Institute for Astronomy, 640 N. Aohoku Place, Hilo, HI 96720, USA}

\author{Larry Denneau}
\affil{University of Hawaii, Institute for Astronomy, 2680 Woodlawn Drive, Honolulu, HI 96822, USA}

\author{Michael Tucker}
\affil{University of Hawaii, Institute for Astronomy, 2680 Woodlawn Drive, Honolulu, HI 96822, USA}

\author{Benjamin J. Shappee}
\affil{University of Hawaii, Institute for Astronomy, 2680 Woodlawn Drive, Honolulu, HI 96822, USA}

\author{Mark E. Huber}
\affil{University of Hawaii, Institute for Astronomy, 2680 Woodlawn Drive, Honolulu, HI 96822, USA}

\author{Anna V. Payne}
\affil{University of Hawaii, Institute for Astronomy, 2680 Woodlawn Drive, Honolulu, HI 96822, USA}

\author{Aaron Do}
\affil{University of Hawaii, Institute for Astronomy, 2680 Woodlawn Drive, Honolulu, HI 96822, USA}

\author{Chien-Cheng Lin}
\affil{University of Hawaii, Institute for Astronomy, 2680 Woodlawn Drive, Honolulu, HI 96822, USA}

\author{Michael S. Connelley}
\affil{University of Hawaii, Institute for Astronomy, 640 N. Aohoku Place, Hilo, HI 96720, USA}

\author{Watson P. Varricatt}
\affil{University of Hawaii, Institute for Astronomy, 640 N. Aohoku Place, Hilo, HI 96720, USA}

\author{John Tonry}
\affil{University of Hawaii, Institute for Astronomy, 2680 Woodlawn Drive, Honolulu, HI 96822, USA}

\author{Kenneth Chambers}
\affil{University of Hawaii, Institute for Astronomy, 2680 Woodlawn Drive, Honolulu, HI 96822, USA}

\author{Eugene Magnier}
\affil{University of Hawaii, Institute for Astronomy, 2680 Woodlawn Drive, Honolulu, HI 96822, USA}



\begin{abstract}

We report photometry and spectroscopy of the outburst of the young stellar object
Gaia19bey.
We have established the outburst light curve
with archival Gaia ``$G$'', ATLAS ``Orange'', ZTF $r$-band and Pan-STARRS ``$rizy$''-filter
photometry, showing an outburst of $\approx$ 4 years duration,
longer than typical EXors but shorter than FUors.
Its pre-outburst SED shows a flat far-infrared spectrum, confirming
the early evolutionary state of Gaia19bey and its similarity to other deeply embedded young stars experiencing outbursts.
A lower limit to the peak outburst luminosity 
is $\approx$ 182 {$L_\odot$} at an assumed distance of 1.4 kpc, the minimum
plausible distance. 
Infrared and optical spectroscopy near maximum light showed an emission line spectrum, including \ion{H}{1} lines,
strong red \ion{Ca}{2} emission, 
other metal emission lines, infrared CO bandhead emission, and a strong
infrared continuum.
Towards the end of the outburst, the emission lines have all but disappeared and the spectrum has changed into
an almost pure continuum spectrum. 
This indicates a cessation of magnetospheric accretion activity.
The near-infrared colors have become redder as Gaia19bey has faded,
indicating a cooling of the continuum component.
Near the end of the outburst, the only remaining strong emission lines are
forbidden shock-excited emission lines.
Adaptive optics integral field spectroscopy shows the H$_2$ 1--0 S(1) emission with
the morphology of an outflow cavity and the extended emission in the [\ion{Fe}{2}] line
at 1644 nm with the morphology of an edge-on disk. 
However, we do not detect any large-scale jet from Gaia19bey.

\end{abstract}





\section{Introduction} \label{sec:intro}

Young Stellar Objects (YSOs) 
in Spectral Energy Distribution (SED) classes I and II, 
i.e. stars in their late accretion phase,
often show substantial
variability due to instabilities in the accretion process
and variations of the extinction.
The accretion characteristics of young stars have recently
been reviewed by \citet{Hartmann.2016.ARA&A.54.135H} and we
follow their general line of discussion and the references therein.

Traditionally, the photometric outbursts of young stars caused by
increased accretion rates were divided by
\citet{Herbig.1977.ApJ.217.693H} into two classes, based on
a small number of photographically discovered cases: 
FU Orionis objects (FUor) and EX Lupi objects (EXor).
The outburst amplitude is similar for both classes, but the
FUor outbursts last for decades to centuries, while
EXor outbursts last from months to maybe a few years.

The first known outburst of a young stellar object, FU Orionis, still remains 
the most substantial of these accretion instability events, having
hardly declined in brightness from its maximum as first discussed by
\citet{Herbig.1977.ApJ.217.693H}. 
For a recent comparison of FUor-type light curves see \citet{Hillenbrand.2019.ApJ.874.82H},
who compare the recently discovered FUor PTF14jg with 
the classical examples, and the comprehensive review of eruptive YSOs by
\citet{Audard.2014.prpl.conf.387A}.

The short duration EXor outbursts are typically repetitive on timescales
of a few years to decades, as illustrated in the case of the
deeply embedded EXor V1647 Ori in a series of papers culminating with
\citet{Aspin.2011.AJ.142.135A},
and more broadly reviewed by \citet{Audard.2014.prpl.conf.387A}.

These two classical types of YSO outburst are also distinct spectroscopically.
EXors are characterized by an emission line
spectrum probably produced in optically thin funnel flows in
a magnetospheric accretion scenario and veiling of photospheric
absorption lines by an UV and optical continuum produced in
high-temperature shocks. In contrast, the more
substantial FUor outbursts show a low-gravity absorption line
spectrum reminiscent of a supergiant photosphere thought to be
caused by a self-luminous optically thick accretion disk.

As more and more YSO outbursts have been observed thanks to
better all-sky monitoring at optical wavelengths, mostly at
``red'' wavelengths, 
and substantial surveys of star-forming regions in the infrared,
these newer discoveries have begun to fill a continuum of 
light curve characteristics such as amplitude, rise time, and rate of decline.
Such outbursts with intermediate characteristics, in particular
those more deeply embedded than EX Lupi itself and associated
with reflection nebulosity or outflow features and generally being more
luminous than the prototypical EX Lupi, have been 
called ``Newest EXors'' by
\citet{Lorenzetti.2012.ApJ.749.188} and
``MNors'' by \citet{ContrerasPena.2017.MNRAS.465.3011C, ContrerasPena.2017.MNRAS.465.3039C}

The spectroscopic properties of eruptive young stars also show
a wide diversity that challenges the simple EXor vs. FUor classification scheme
established by \citet{Herbig.1977.ApJ.217.693H}. In particular, some objects show a nearly
featureless continuum spectrum, indicative of dominant dust emission,
and therefore cannot be classified as either FUor or EXor.
The first of these objects, initially called a ``deeply embedded outburst star'' was OO~Ser discovered and studied by
\citet{1996ApJ...468..861H, 2012ApJ...744...56H} that
also exhibited an outburst duration of about two decades, longer
than any EXor and shorter than any FUor. 

In this paper, we use the term ``EXor'' in the broad sense of a YSO
outburst of short duration of at most a few years and with an emission line spectrum.
We do not imply that all EXors must closely resemble the prototypical
EX Lupi, a low-luminosity (L=0.73 {$L_\odot$}) T Tauri star (SED Class II) without associated nebulosity
and essentially no extinction, with a rich optical emission line
spectrum
\citep{Herbig.2007.AJ.133.2679.EXLupi}
but a near-infrared absorption spectrum \citep{Sipos.2009.A&A.507.881.EXLupi}.

The total number of known YSO outburst objects is still quite small,
at most a few dozen objects of the FUor, EXor, and intermediate classes combined.
A detailed study of each individual case is therefore important for
understanding the full range of eruptive phenomena displayed in this type
of objects.

We report here on the preliminary characterization of Gaia19bey at
$\alpha$ = 20:40:44.39 $\delta$ = +46:53:21.34 (J2000.0), first noted in a Gaia alert (Gaia19bey)
\citep{Gaia-2016A&A...595A...1G} on 2019 April 3 as
 ``red source brightens by 3 mags over 2.5 years''.
\footnote[1] {http://gsaweb.ast.cam.ac.uk/alerts/alert/Gaia19bey/}.

\section{Gaia19bey: Location and Distance}

Except for its inclusion in the IRAS and 2MASS point
source catalogs as IRAS 20390+4642 and \\
J20404439+4653215, respectively, 
Gaia19bey has not been studied before.
It is not included in the Gaia-DR2 catalog 
\citep{Gaia.2018.A&A.616A.1G},
presumably because
it was too faint when those data were taken.
Consequently, we do not know its distance directly.

Gaia19bey is located at
Galactic coordinates l = 85.48489\arcdeg b = +3.07714\arcdeg between 
the outer perimeter of the high extinction region
Cygnus-X centered on Cygnus OB2 and the Cygnus OB7 region, summarized in
the review chapter on Cygnus in \citet{Reipurth.2008.hsf1.book.36R}.
The Cygnus-X star forming complex is at a distance of 1.40 $\pm$ 0.08 kpc
based on measurements of embedded methanol and water masers by 
\citet{Rygl.2012.AA.539A.79.CygX.distance}.
However, in this region of the sky, distance measurements are notoriously difficult 
since the line of sight is basically
along the local spiral arm, as 
\citet{Rygl.2012.AA.539A.79.CygX.distance}
had pointed out, and objects at different distances may
be superposed on the sky.

\begin{deluxetable}{ccccc}
\tabletypesize{\scriptsize}
\tablecaption{Gaia Parallaxes}
\tablewidth{0pt}
\tablehead{
\colhead{Gaia ID} & \colhead{Parallax [mas]} & \colhead{Error [mas]} & \colhead{RUWE} &\colhead{Dist [kpc]}
}
\startdata
2167375909294267392 & 0.449 & 0.149 & 0.944 & 2.2 \\
2167375913600701568 & 0.729 & 0.035 & 0.985 & 1.4 \\
2167375707442271744 & 0.356 & 0.103 & 0.969 & 2.8 \\
2167375913600703744 & 0.223 & 0.061 & 0.996 & 4.5 \\
2167375707442274304 & 0.369 & 0.083 & 1.001 & 2.7 \\
2167375909294286208 & 0.707 & 0.273 & 1.065 & 1.4 \\
\enddata
\end{deluxetable}

A small number of low-extinction stars near this group of reddened objects
around Gaia19bey have Gaia DR-2 \citep{Gaia.2018.A&A.616A.1G} distances with better than 3$\sigma$ significance.
These are indicated as yellow labels (in kpc) in Figure~1 
and are listed in Table 1 in order of increasing RA. All stars listed there have Renormalised Unit Weight Error (RUWE) around 1.0,
well below the upper limit of 1.4 considered to be an indicator of astrometrically reliable data
\citep{Lindegren.2018.AA.616.2.RUWE}.
The two shortest distances
among this group of 5 stars are consistent with the distance to Cygnus-X of 1.4 kpc.
However, three of these stars have a distance in the range of 2.2 - 2.8 kpc, and
one is measured at a distance of 4.5 kpc. 
From this small number of Gaia distances, we conclude that the
distance to the Cygnus-X region of 1.4 kpc is a lower limit
to the distance to Gaia19bey, but that its distance is probably
higher, at least in the range above 2 kpc. If we 
take the single 4.5 kpc measurement of an unreddened star 
as reliable,
this makes 4.5 kpc a lower limit to its distance.
For the purpose of determining the luminosity of the object,
we use the lower limit of 1.4 kpc as the distance to
Gaia19bey.

We have inspected publicly available images from the Pan-STARRS 3-$\pi$ survey
\citep{Chambers.2016.arXiv.161205560C},
the INT Photometric H-Alpha Survey IPHAS
\citep{Drew.2005.MNRAS.362.753.IPHAS},
the Two Micron All-Sky Survey 2MASS \newline
\citep{Skrutskie.2006.AJ.131.1163},
the United Kingdom Infrared Deep Sky Survey UKIDSS \citep{Lawrence.2007.MNRAS.379.1599.UKIDSS}
and the Wide-Field Infrared Survey Explorer WISE
\citep{Wright.2010.AJ.140.1868.WISE}.
A subset of these archival images has been included in the three
color composites in Figure~1. In all of these images, Gaia19bey is
the star in the center of the image, very faint at optical wavelengths,
but dominant in the far infrared. The middle panel shows the Pan-STARRS PS1 z-band
\citep{Tonry.2012.ApJ.750.99.PS1-photometric-system}
and the UKIDSS $J$ and $K$ band images, showing that Gaia19bey is the brightest
in a small group of highly reddened stars clearly distinct from the
surrounding star field. We work under the assumption that this group of
reddened stars outlines a small molecular cloud seen in absorption against
more distant stars in the local spiral arm. 
The bottom panel of Figure~1 is composed of WISE 3.4 $\mu$m, WISE 12 $\mu$m and
AKARI Far-Infrared Surveyor (FIS) Wide-S (90 $\mu$m)
\citep{Murakami.2007.PASJ.59S.369.AKARI}
data and shows that Gaia19bey is the
only object in this small group with 
strong mid-IR and far-IR emission and that it does not suffer confusion with other
far-IR point sources so that its SED can be reliably measured.

The archival PS1 3$\pi$ survey images used in Fig.~1
are composites of observations taken between MJD 55000 and 56500 and represent
\footnote[2]{MJD = JD - 2400000.5}
what we consider the quiescent state of Gaia19bey, despite some relatively minor variations
that will be discussed in section 4.1. 
The UKIDSS and WISE images used in Figure~1 were taken
in the same quiescent phase of the light curve as indicated in the top panel of Figure~2. 
The only image used in Fig.~1 that we cannot put into the context of the light curve
is the 90 $\mu$m images from AKARI.
The AKARI mission 
\citep{Murakami.2007.PASJ.59S.369.AKARI}
operated between MJD 53863 and
54338, a few years before Pan-STARRS started its observations.

While the whole Cygnus-X region shows numerous H$\alpha$ emission regions,
a careful inspection of the available optical and near-infrared images,
in particular the H$\alpha$ images of the IPHAS survey
\citep{Drew.2005.MNRAS.362.753.IPHAS, Barentsen.2014.MNRAS.444.3230.IPHAS}
did not show any reflection nebulosity or large scale jet-like features
associated with Gaia19bey at seeing-limited resolutions.

\begin{figure}[h]
\begin{center}
\includegraphics[angle=0.,scale=0.38]{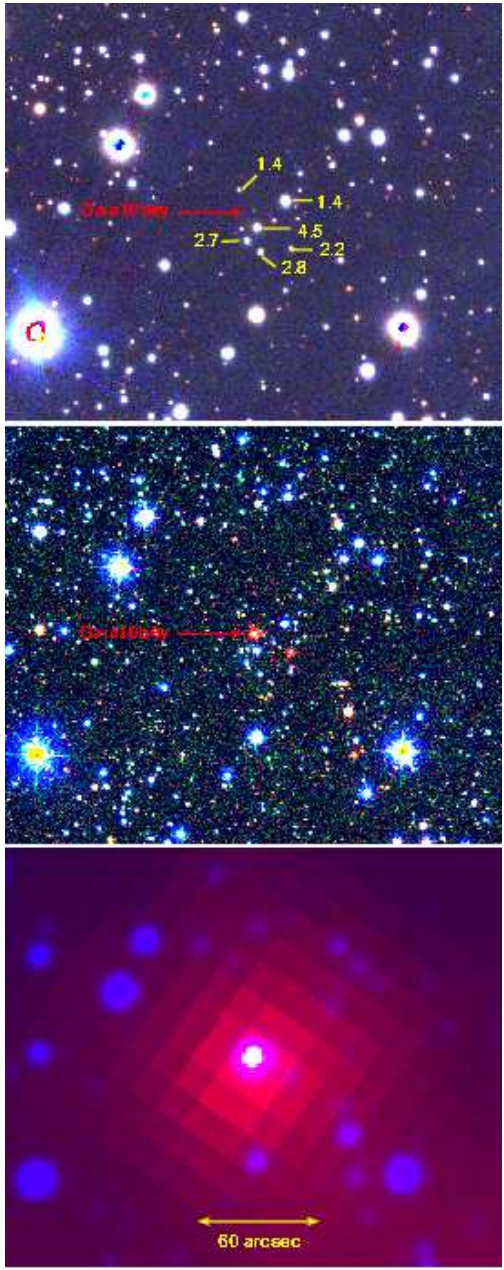}
\caption{
Color composite images of the Gaia19bey region.
The top panel shows PS1-g, PS1-r, and PS1-i images and shows
mostly the low-extinction objects in this field. We have labeled some optically detectable stars
near the position of Gaia19bey with their distance in kpc from the
Gaia DR-2. 
The middle panel shows PS1-z, UKIDSS-$J$, and UKIDSS-$K$. 
The bottom panel shows WISE 3.4 $\mu$m, WISE 12 $\mu$m, and AKARI Wide-S 90 $\mu$m.
Gaia19bey is the only object with strong 12 $\mu$m flux
and is the only FIR point source.
All the data shown in this figure represent the
quiescent state of Gaia19bey, to the best of our knowledge of
its photometric history, 
except the AKARI Wide-S 90 $\mu$m, for
which we cannot establish the context in the light curve.
}
\end{center}
\end{figure}

\section{Observations and Results}

\subsection{Gaia and ATLAS}
The Gaia $G$-band photometric data were obtained from 
the Gaia alert data base and form the basis for the light curve
in Figure~2.
\footnote[3]{\url{http://gsaweb.ast.cam.ac.uk/alerts}}
From the start of the measurements on 2014 November 30 to 2016 August 4, Gaia
did not return significant detections of our object, and from the faintest reported
measurements we treat these non-detections as upper limits at $G$ = 20.5.

Many of the bright data points of the optical light curve of Gaia19bey are archival data from the
ATLAS project described by \citet{Tonry.2018PASP..130f4505T}. 
ATLAS usually takes more than one image of any given region of the sky in each suitable
night to follow fast-moving asteroids. For the light curve points in Figure~2, 
we have median-combined the individual measurements
for each night when the photometric zero points and sky brightness were stable.
From the small number of epochs when closely coinciding Gaia and ATLAS or Pan-STARRS
data points were available, the magnitude offset in each filter was determined
and the other light curve points in that filter were shifted accordingly, so
that the full light curve is effectively in Gaia-$G$ magnitudes.
We did not have the data to also match second order (color) effects.
The ATLAS data points have fairly large errors in the fainter phases of the
light curve and we therefore include only the brighter (effectively $G <$ 18.0) 
ATLAS ``Orange'' data points in Figure 2.
The current outburst is the largest observed in the past decade, but smaller
variations are seen in the older Pan-STARRS data.

\begin{figure}[h]
\begin{center}
\includegraphics[angle=0.,scale=0.52]{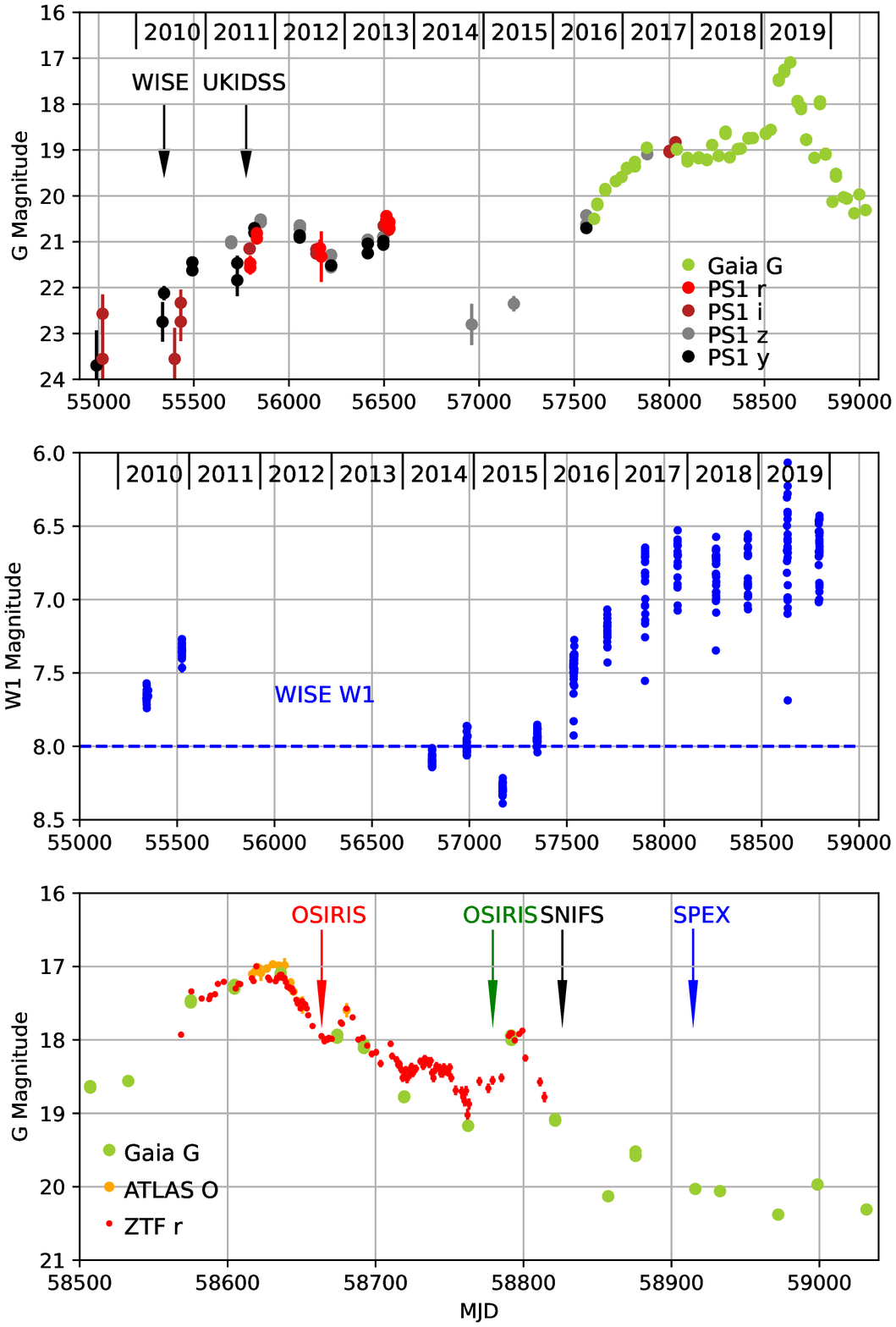}
\caption{
Top Panel:
Light curve of Gaia19bey from Pan-STARRS r,i,z, and y-band photometry,
Gaia photometry (``G''-filter), and ATLAS ``$O$'' data.
Two Pan-STARRS data points around MJD 57000 indicate a minimum just prior to the rise
towards the present maximum.\\
Middle Panel:
WISE W1 photometry, combining data from the cryogenic mission, NEOWISE, and
the resumed NEOWISE. The W1 band saturation limit is indicated by a dashed line.\\
Bottom Panel:
The arrow symbols indicate the epochs when spectra were taken,
and the colors are the same as used in Figures 5 (Section 3.4) and Figure 6 (Section 3.5).
}
\end{center}
\end{figure}

\subsection{Pan-STARRS-1}
The Panoramic Survey Telescope and Rapid Response System 
\citep[Pan-STARRS]{Kaiser.2010.SPIE.7733E.0EK}
is a wide-field optical imaging
sky survey project.  The PS1 3$\pi$ survey 
\citep{Chambers.2016.arXiv.161205560C}
patrolled the entire sky north of Declination.$=-30^\circ$ at multiple epochs between June 2009
and April 2014 and in 5 filters (\ensuremath{grizy_{\rm P1}})
\citep{Tonry.2012.ApJ.750.99.PS1-photometric-system}.
The 5$\sigma$ single-epoch depths are down to 22.0,
21.8, 21.5, 20.9, and 19.7~mag, respectively and photometric calibration uncertainty
of the survey is $\sim0.01$~mag.
We extracted the
photometric data from the PS1 third full reduction data set through the PS1 internal
database system.
Additionally, we ingested some post-PS1 3$\pi$ survey data covering
our target, hence the time baseline extends up to May 2017.  The additional data
were also calibrated by a reference catalog generated from the previous reductions
data 
\citep{Magnier.2016arXiv161205242M}.
For the light curve in Figure~2, we have shifted the magnitudes in any of the
PS1 filters to the Gaia $G$ photometry, when closely coincident epochs were available.
This shifting is just the first order
of adjusting the different filter data and does not include color correction because
coincident color information was not available.
These adjusted Pan-STARRS data give an approximate light curve covering the years from 2009 to 2017 effectively
in Gaia $G$ magnitudes,
when neither Gaia nor ATLAS data are available. 

\subsection{Archival Data}

To complement the ATLAS data, we have downloaded 
\footnote[4] {https://irsa.ipac.caltech.edu}
archival
$r$-band photometry of Gaia19bey from the Zwicky Transient Facility (ZTF)
\citep{Bellm.2019.PASP.131.8002.ZTF}
archive \citep{Masci.2019.PASP.131.8003.ZTFarchive}.
These data reach about one magnitude fainter than the ATLAS
data and cover the outburst maximum very well. Similar to the
procedure for ATLAS and Pan-STARRS, we shifted the ZTF magnitudes
by a fixed amount to match the Gaia $G$ photometry.

We have included archival data from the WISE 
\citep{Wright.2010.AJ.140.1868.WISE}
and 
NEOWISE 
\citep{Mainzer.2014.ApJ.792.30.NEOWISE}
missions
downloaded from the NASA/IPAC Infrared Science Archive $^4$
in the light curve (Middle panel in Figure 2). Most of the WISE data in the W1 and W2 bands on Gaia19bey are 
saturated, but the fainter W1 data points prior to the present outburst are unsaturated
and confirm the brightness minimum that is otherwise only indicated by
just two Pan-STARRS measurements. The W1 band data points above saturation still
contain some useful information in that they confirm the basic shape of the
light curve during the outburst.

Figure~3 shows the spectral energy distribution (SED) of Gaia19bey
based on ground-based and space-based survey data obtained from
VizieR.
The infrared and far-infrared
data are from the compilation of catalog data by \citet[and references therein]{Abrahamyan.2015.AC.10.99A}
and include data from WISE \citep{Wright.2010.AJ.140.1868.WISE}, MSX \citep{Egan.1996.AJ.112.2862.MSX},
AKARI \citep{Murakami.2007.PASJ.59S.369.AKARI},
and IRAS \citep{Neugebauer.1984.ApJ.278L.1.IRAS}.
The 2MASS data are from the point source catalog \citep{Skrutskie.2006.AJ.131.1163}.
The photographic data points (DSS) are from the guide-star catalog of
\citet{Lasker.2008.AJ.136.735L}.
In the optical and near-infrared we have compiled all available 
individual photometric values from Pan-STARRS PS1,
the 2MASS and UKIDSS surveys, and NEOWISE.
All these data were taken without any knowledge of the light
variations of Gaia19bey and do not represent the full range of 
brightness variations that Gaia19bey may have experienced in
the past, but they give an impression of the typical variations of
the short end of the SED with its variability.
Since Gaia has covered this object from invisibility, i.e. G $>$ 20.5, early
in its mission to the current maximum, those data points
represent a good lower limit of the total variability this
object has experienced in the past.
Also, we note that Gaia19bey must have been
in a previous, but unrecognized outburst when the 2MASS photometry was obtained on 1998 November 3
since the 2MASS catalog photometry is brighter than the maximum
of infrared photometry measured during
the present outburst.
For wavelength longer than 4.6 $\mu$m, we do not have 
multi-epoch data from individual missions, but some
of the observed scatter may still be due to variability.

\begin{figure}[h]
\begin{center}
\includegraphics[angle=0.,scale=0.46]{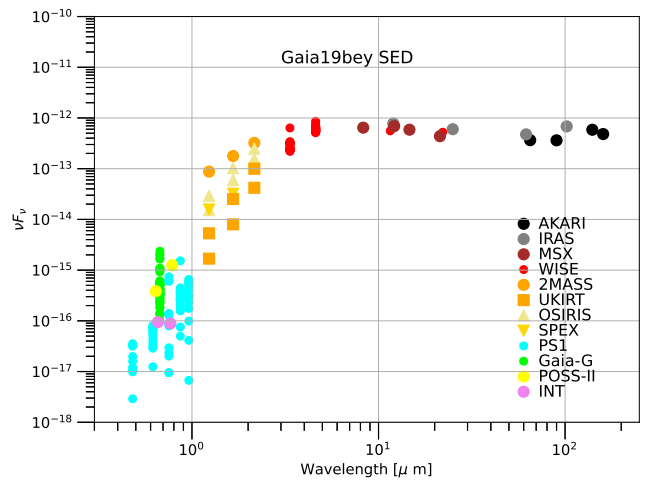}
\caption{
Spectral energy distribution (SED) of Gaia19bey
from catalog data available
in the VizieR data base and Pan-STARRS, Gaia, and NEOWISE multi-epoch
photometry. Individual references are in the text.
The SED longward of
the WISE 4.6 $\mu$m band is remarkably flat.
We have included the Gaia, Pan-STARRS, $J$, $H$, and $K_s$ photometry reported here to
give an impression of the near-infrared variability.
The 2MASS photometry represents the brightness in an earlier
outburst that went unnoticed.
}
\end{center}
\end{figure}

\subsection{UKIRT WFCAM Imaging}
Deep infrared images
were obtained in the $J$, $H$, $K$ bands and the narrow-band S(1) line filter with the Wide-Field Camera (WFCAM) described by \citet{Casali2007} on UKIRT
on 2020 June 20.
The exposure time of individual frames was 1.0 s in J, H, and K with 5 coadds per position and a total of 72 dithered
frames taken, resulting in a total integration time of 360s.
We have verified that the detector did not saturate in the K band on this relatively
bright object.
In the S(1) filter, the individual frame exposure time was 20s, and the total integration
time in the 72-point dither pattern was 1440 s.
The continuum subtracted image in Figure~4 is the S(1) line image with 1.2 times the K-band
in the 1 s individual exposures
image subtracted for optimal subtraction of field star. The stars in the group of highly reddened
stars near Gaia19bey had higher $K$-band flux than the average field stars and are therefore
oversubtracted in Figure~4.
It is noteworthy that the UKIDSS image of Gaia19bey had an exposure time of 5 sec
and was in the non-linear regime of the detector, despite the fainter magnitude at the time. 
\footnote[4] {http://casu.ast.cam.ac.uk/surveys-projects/wfcam/technical/linearity}
To account for this possible systematic effect, we estimate a larger error for this data point, as indicated in
Table 2.

\begin{figure}[h]
\begin{center}
\includegraphics[angle=0.,scale=0.50]{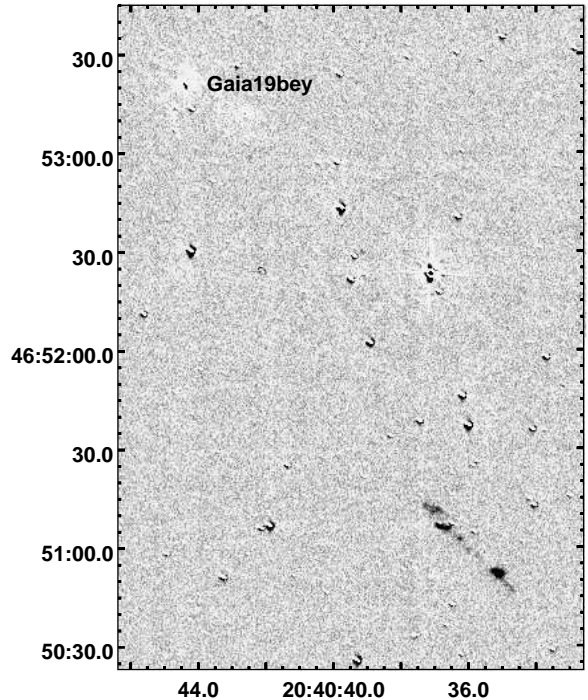}
\caption{
Continuum-subtracted UKIRT/WFCAM image of the wider area aroung Gaia19bey in the H$_2$ 1--0 S(1) emission line.
The H$_2$ jet shown here is probably not emerging from Gaia19bey since the shape of the bow-shocks suggests
that it moves from the SW to the NE.
}
\end{center}
\end{figure}

\subsection{UH 88" SNIFS Optical Spectroscopy}

We obtained an optical spectrum of Gaia19bey with the
``SuperNova Integral Field Spectrograph'' (SNIFS)
described in \citet{Lantz.2004.SPIE.5249.146L}
at the UH 88" telescope on 2019 December 9 (MJD 58826).
SNIFS is a dual-channel instrument, but we only show
the red part of the spectrum in Figure~5 because the
blue channel spectrum was too faint to show any features. The spectral
resolution is approximately R $\approx$ 1300.
The spectrum has not been corrected for
telluric absorption, and the absorption feature of O$_2$ at
$\approx$ 760 nm 
\citep{Rudolf.2016.AA.585.113}
is clearly visible, but does not affect
the interpretation of this spectrum.
The line identifications are based on the study of Herbig AeBe stars
by \citet{Hamann.1992.ApJS.82.285.CaIIHerbigAeBe}.

\begin{figure}[h]
\begin{center}
\includegraphics[angle=0.,scale=0.31]{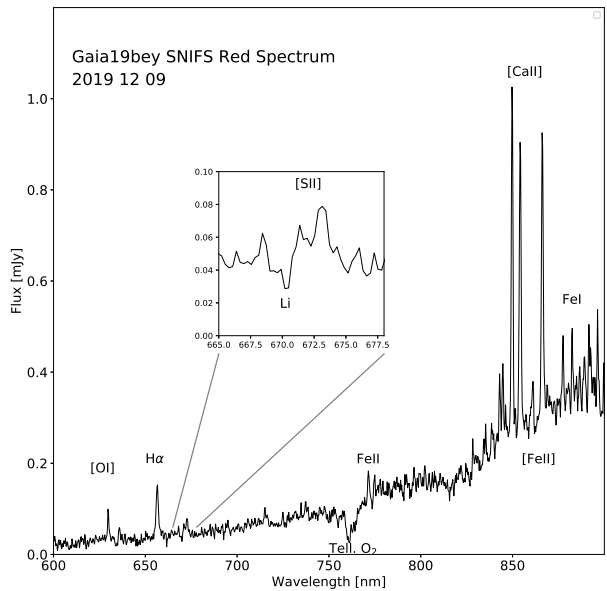}
\caption{
Optical (red) spectrum of Gaia19bey 
obtained with the SNIFS integral-field spectrograph (R $\approx$ 1300) at the UH 88-inch Telescope. 
The main emission lines are labelled in the Figure;
the most prominent emission line system is the \ion{Ca}{2} triplet.
The magnified insert shows the weak detection of the \ion{Li}{1} line
at 670.8 nm, the only absorption line detected in the spectrum.
}
\end{center}
\end{figure}

\subsection{Infrared Spectroscopy}

We obtained infrared spectroscopy covering parts of the
wavelength range 1.0 - 2.5 $\mu$m during
two observing nights at the Keck 1 telescope,
and using director's discretionary time on the NASA
Infrared Telescope Facility (IRTF).
All infrared spectra are shown in Figure~6.

\subsubsection{Keck OSIRIS Infrared Spectroscopy}

Gaia19bey was observed using the OSIRIS integral field
spectrograph 
\citep{Larkin2006SPIE.6269E..1AL}
at the Keck 1 telescope on 2019 June 29 and 2019 October 23.
The median spectral resolution
at the wavelength of our H$_2$ 1--0 S(1) line observations
is R $\approx$~3500. 
The night of 2019 June 29 was mostly cloudy and Gaia19bey
was observed as a backup program towards the end of the night
through gradually clearing cirrus clouds. With OSIRIS, the three
order sorting filters $Jbb$, $Hbb$, and $Kbb$ were used, and
observations of the
telluric absorption standard HD 199217 were interleaved
with the observations of Gaia19bey.  
While the correction of telluric atmospheric absorption
worked reasonably well under these unstable weather conditions,
the flux calibration is not reliable on the level of
a few percent. Further, the large number of emission lines,
in particular all the hydrogen lines in the $Hbb$ filter,
made fitting of the continuum flux difficult.

The spectrum of a star was reduced
by first extracting the spectra of each spaxel (lenslet)
using the OSIRIS data reduction pipeline (DRP).
In order to subtract the night sky spectrum, 
aperture photometry with sky subtraction was done on the individual planes of the data cubes
produced by the DRP, using a custom IRAF script based on the IRAF task apphot
\citep{Tody1986}.
The resulting spectrum was reformatted as input
into the Spextool Xtelcorr tool developed by
\citet{Cushing.2004.PASP.116.362.SPEXTOOL}
for the
IRTF SPEX instrument, so that we have commonality in the
telluric correction and flux calibration procedures
for the spectra from Keck and IRTF, which will be
described further in section 3.6.2. 
 
Some of the OSIRIS data cubes were used to extract continuum
subtracted images (Figure~7) of Gaia19bey in the emission lines
of [\ion{Fe}{2}] at 1644 nm and H$_2$ 1--0 S(1).
The Br$\gamma$, \ion{Na}{1}, and the CO bandhead
emission lines originate in an unresolved region around the star.
While most spectra were obtained with the 50 mas spatial scale and the $Jbb$, $Hbb$, and $Kbb$ filters of the OSIRIS
instrument, a few setup images were taken with the wider 100 mas scale with the $Kn2$ filter that leads to much wider
spatial coverage of the data cubes. 
In the continuum-subtracted data cube planes, the H$_2$ 1--0 S(1) line emission is spatially resolved and suggests that
it originates from the inner wall of an outflow cavity, as will be discussed in section 4.5.

\subsubsection{IRTF SPEX Infrared Spectroscopy}
The IRTF SPEX instrument
\citep{Rayner.2003.PASP.115.362}
in its short-wavelength cross-dispersed (SXD) mode
was used on 2020 March 6 (UT) 
to obtain a spectrum of Gaia19bey at the faintest flux levels
recorded so far during the present outburst, but still slightly
above the pre-outburst upper limits from Gaia.
We used a 0\farcs3 slit and observed
the A0V star
HD 192538 at airmasses closely similar to that of the object
observations
as the telluric absorption standard.
The Spextool software
was used for the extraction of the spectrum, telluric absorption correction, in particular
for the fitting of the deep hydrogen absorption lines in
the standard star spectrum, and photometric calibration, using the same
standard star for both telluric absorption correction and flux calibration
as described in detail by  
\citet{Cushing.2004.PASP.116.362.SPEXTOOL}.


\begin{figure*}[h]
\begin{center}
\includegraphics[angle=0.,scale=0.60]{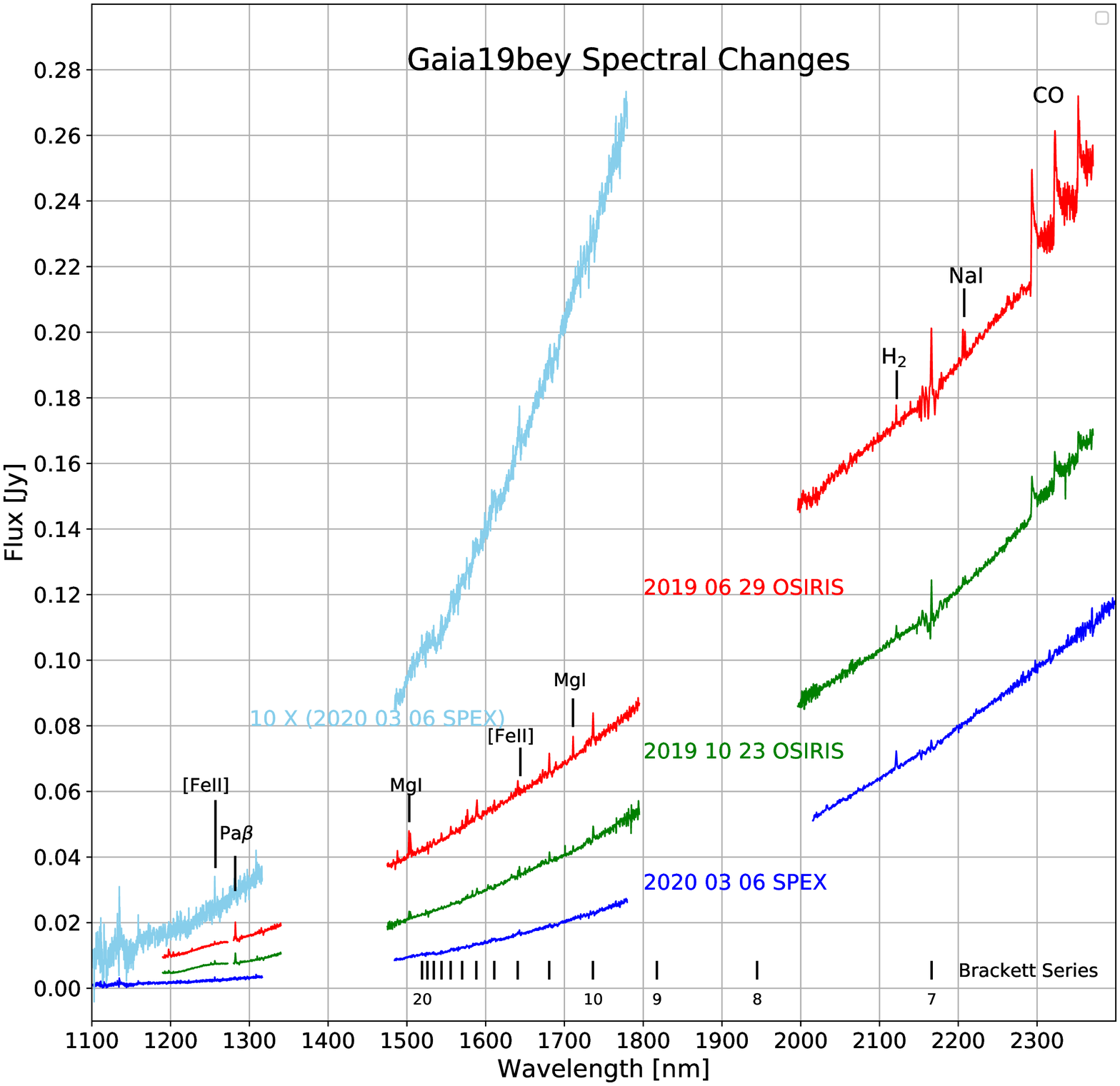}
\caption{
Keck OSIRIS spectra obtained at two different epochs and the IRTF SPEX spectrum.\\
The spectra change from an emission line spectrum near the maximum brightness (red), typical of a deeply embedded EXor outburst, through diminished emission
line strength further into the decline (green) to a pure continuum spectrum (blue) late in the outburst.
The last spectrum (taken with SPEX) shows only shock-excited forbidden lines of [\ion{Fe}{2}] and
H$_2$ probably originating in outflow shock fronts.
}
\end{center}
\end{figure*}

\begin{figure}[h]
\begin{center}
\includegraphics[angle=0.,scale=0.50]{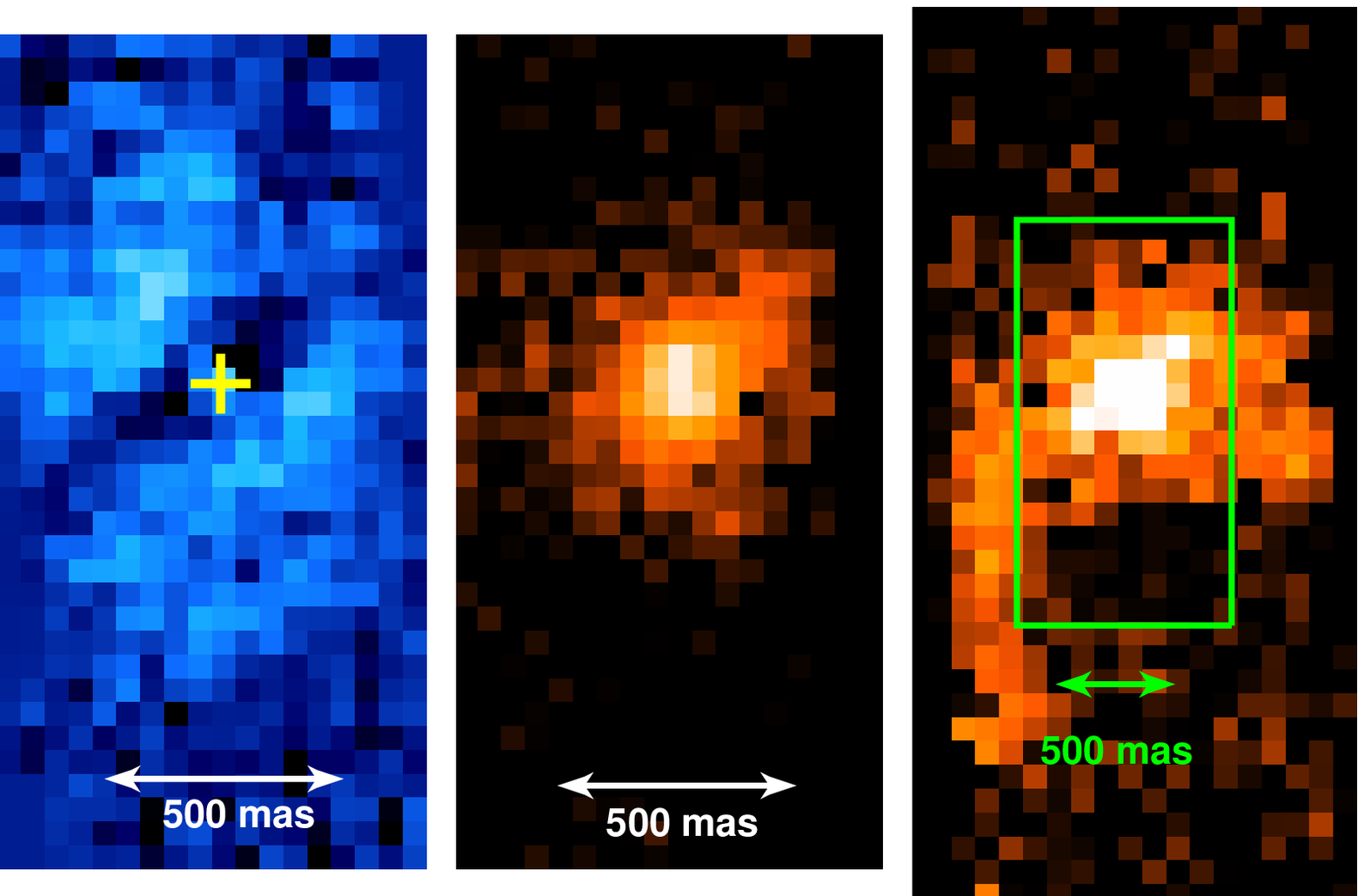}
\caption{
The left panel, colored in blue, shows the continuum subtracted emission in the 1644 nm line of [\ion{Fe}{2}] at a spaxel scale
of 50 mas.
The central panel shows the continuum subtracted H$_2$ 1--0 S(1) emission at 50 mas per pixel.
The left and center panel have a field of 0\farcs9 $\times$ 1\farcs8.
The right panel shows the same H$_2$ emission taken with the 100 mas spaxel scale of OSIRIS and has a field of 1\farcs9 $\times$ 3\farcs6, showing at least
part of the outline of an outflow cavity.
}
\end{center}
\end{figure}

\subsubsection{Photometry Extracted from the Spectra}

We have extracted photometry from the flux calibrated OSIRIS and SPEX spectra.
The OSIRIS spectra did not cover the full wavelength range of the 2MASS filters
\citep{Skrutskie.2006.AJ.131.1163}
so the best method that we could use consistently for the OSIRIS and SPEX spectra
was to use a line-free average of the measured flux density near the effective wavelength
of the 2MASS filters as an approximation of 2MASS photometry. 
While the OSIRIS spectra
are of high signal-to-noise ratio, the main uncertainties in these synthetic
magnitudes stem from systematic calibration issues. 
The 2MASS catalog magnitude of HD199217 has errors of 0.029 mag in J, 0.051 mag in H, and
0.024 mag in K$_s$. The difference in
airmass between the object and standard star observations of maximally
0.05 airmass, combined with an estimate of the extinction coefficient
under the prevailing weather conditions of 0.044 mag/airmass from
\citet{Tokunaga.2002.PASP.114.180.MKO.filters}
give only 0.002 mag uncertainty and is negligible compared to the
magnitude errors of the calibration star itself.
Other erors may come from different Strehl ratio achieved by the
adative optics system between the object (laser guide star use) and
the standard (natural guide star used). Finally, the night of 
2019 June 29 (UT) was non-photometric with the Gaia19bey data being obtained
in the clearing phase after the passage of a large cirrus band. We estimate that
photometric errors may have been as much as 0.05 mag rms in that night,
while we expect the nominally photometric night of 2019 October 23 to
have allowed 0.03 magnitudes photometric errors.
We are including these error estimates in Table 2.
The absolute calibration problems also make the Keck OSIRIS spectra
problematic for any attempt to fit models of the continuum emission.
We will therefore refrain from discussing possible modeling of
this continuum emission.

In contrast to the situation with an integral field spectrograph like
OSIRIS, the SPEX data were obtained with a slit of 0\farcs3 width,
chosen to maximize the spectral resolution, but clearly not accepting
all the light from the object and therefore problematic for absolute
flux determination. While calibrating the spectra with a standard star can give
absolute photometry, variations in the seeing are the dominant limitation to the photometric
precision of this technique. We have obtained 4 individual exposures of the
telluric absorption standard HD192538 and 16 exposures of Gaia19bey.
For each individual exposure, the signal profile along the slit direction
was fitted with a Gaussian function, and the standard deviation of the
best fit signals was computed.
We have added the standard deviations of the standard star and the object 
in quadrature to arrive at an
estimate for the error of the measured flux levels of 19\% or 0.2 mag.
Despite these limitations, the photometry derived from the SPEX data
is valuable as it provides absolute fluxes simultaneous with the
emission line measurements.
We also note that the SPEX photometry where the full spectral range
is recorded simultaneously, has much smaller errors in the color
of the object, since the seeing, the pointing errors, as
well as possible absorption by cirrus clouds, are very similar in
adjacent photometric bandpasses, so that these effects largely
subtract out.

\section {Discussion}

Our goal is to put the newly discovered and on-going outburst of
Gaia19bey in the context of
other, better studied eruptive events in young stars.
As already mentioned in the Introduction, we will use the term ``EXor'' in the broadest sense describing
a young eruptive variable with an emission-line spectrum and an outburst
duration of at most a few years.

\subsection{Light Curve}

The light curve (Figure~2) shows a broad maximum from MJD 57500, when Gaia began
recording significant detections, to MJD 59000, when the light curve
appears to have faded back to within half a magnitude of the pre-outburst levels below $G$ = 20.5. 
The light curve showed a brief spike up to the maximum of $G$ = 17.09
on 2019 June 1 (MJD 58605) that had a total duration of less than one
year and led to the Gaia alert at a magnitude of $G$ = 17.49.
While the one-year brightness spike is of the same duration as typical EXor outbursts, 
the overall outburst duration above $G$ = 20.5 (the Gaia detection limit) of $\approx$ 4 yr is
longer than those found in classical EXors, but much shorter than that of
FUors. The longer duration is in line with many of the higher luminosity eruptive events
discovered and studied at infrared wavelengths by
\citet{Lorenzetti.2012.ApJ.749.188}, who called them ``Newest EXors'' and that
\citet{ContrerasPena.2017.MNRAS.465.3039C} used to tentatively define
the new MNOR class.

Continguous coverage of the light curve started around MJD 55000 when Pan-STARRS 1 (PS1)
began the PS1 3$\pi$ sky survey. Until the Gaia mission started reporting significant
detections of Gaia19bey, these 
and a few unsaturated $W1$ data points from the WISE and NEOWISE missions
are the only available observations of this object.
The earliest Pan-STARRS measurements, combined from all the filters, between MJD 55000 and 55300 ($\approx$2009) showed
a brightness near $G$ = 23. Around MJD 55500, the brightness increased by 2 magnitudes to
$\approx$21 mag, and this rise was partly confirmed by WISE.
The light curve between MJD 56500 and the
start of the Gaia data at MJD 57500 is only poorly covered since PS1 had completed the
3$\pi$ sky survey and was now concentrating on asteroid searches and only covered
the position of Gaia19bey on two occasions. 
This flux minimum was, however, observed by the NEOWISE-R mission with four
additional data points. These data points were, in fact, the only NEOWISE $W1$ data
points that were not saturated and therefore are a reliable confirmation of
the minimum prior to the outburst when the brightness dropped again to $\approx$23 mag. 
In the brighter phase at approximately 21 magnitude,
the object showed additional variations of about one magnitude and timescale
of $\approx$ 500 days. 
The outburst amplitude depends on how the quiescent brightness is defined.
Prior to the present outburst, our Pan-STARRS photometry suggests a minimum of $G$ $\approx$ 22.5 around
MJD 57000,
even though that time interval is poorly covered with only two Pan-STARRS observations. Relative to this minimum,
the outburst amplitude is 5.5 mag in the $G$ band.

We have no spectroscopic data from that time period and from optical photometry
alone cannot distinguish variations due to accretion variations from variations
causes by variable extinction in the light path. Given that the morphology of
the forbidden line emission (Figure~7) strongly suggests a disk seen nearly edge-on, variability
due to extinction variations is very likely to occur and may be the dominant reason
for those brightness variations.
Between the start of the PS1 observations and the first Gaia detections, the data
indicate initially a faint phase, a brighter phase of 1000 days, and again fainter
magnitudes. This is, of course, insufficient to determine that those changes are
periodic. However, if we assume some semi-periodicity, it is likely that the
latest, much brighter outburst coincided with a period of lower extinction, even
though it was not caused solely by lower extinction.
In this working assumption of a combination of extinction variations and a stronger
variation of accretion, we assume that the baseline for the accretion outburst is
the brighter level of the extinction variations in the light curve at $\approx$ 21 mag.
Calculating the outburst relative to this mean 
gives an outburst amplitude of $\approx$ 4 mag, well in the range of other
eruptive events in YSOs. 


\begin{deluxetable}{lrrrc}
\tabletypesize{\scriptsize}
\tablecaption{Near-Infrared Photometry}
\tablewidth{0pt}
\tablehead{
\colhead{Date} & \colhead{J} & \colhead{H} & \colhead{K} &\colhead{Source}
}
\startdata
1998 11 03 & 11.61$\pm$0.02 & 10.05$\pm$0.02 &  8.64$\pm$0.02 & 2MASS \\
2011 08 02 & 15.90$\pm$0.01 & 13.42$\pm$0.01 & 10.86$\pm$0.03 & UKIDSS \\
2019 06 29 & 12.80$\pm$0.06 & 10.65$\pm$0.07 &  8.93$\pm$0.06 & OSIRIS \\
2019 10 23 & 13.51$\pm$0.05 & 11.24$\pm$0.06 &  9.43$\pm$0.05 & OSIRIS \\
2020 03 06 & 14.71$\pm$0.20 & 11.91$\pm$0.20 &  9.90$\pm$0.20 & SPEX \\
2020 06 20 & 14.65$\pm$0.01 & 12.17$\pm$0.01 &  9.92$\pm$0.01 & WFCAM \\
\enddata
\end{deluxetable}

Some evidence
for a prior major outburst comes from the 2MASS catalog 
\citep{Skrutskie.2006.AJ.131.1163}
that shows
$J$, $H$, and $K_s$ magnitudes (Table 1)
brighter than those seen in the present
outburst, and from the fact that the UKIDSS survey recorded much
fainter 
$J$, $H$, and $K$
values that we assume represent the quiescent state of Gaia19bey.
The UKIDSS $K$-band data point is brighter than the saturation limit
reported by \citet{Lucas.2008.MNRAS.391.136.UKIDSS.GPS} and is therefore
not reliable.

While the early parts of the Gaia light curve prior to MJD 58200 appear
to be smooth, starting shortly before MJD 58300, through the maximum, and in the
declining phase we see several brief spikes in brightness
with typical duration of $\approx$ 10 to 50 days and $G$ amplitudes of order 0.5 mag,
the earlier of which are very well confirmed by the high-cadence ZTF observations. 
The durations and amplitudes of these brief spikes are similar to those
reported by \citet{Semkov.2006.IBVS.5683.1.V1647LC, Semkov.2012.IBVS.6025.1.V1647LC}
in the declining phases of the recent outbursts of V1647 Ori.
Since these minor spikes in brightness occur in the phase where our spectra
indicate on-going accretion activity, they are most likely associated with
fluctuations in the magnetospheric accretion process.

\subsection{The SED Variations and Luminosity}

The SED is a typical ``flat spectrum'' distribution
\citep{Greene.1994ApJ...434..614G}. The flat part of the SED extends from
the WISE 2 band (4.6 $\mu$m) out to 160 $\mu$m from AKARI, characterizing Gaia19bey
as a YSO in transition from a deeply embedded infrared object (Class I)
to a moderately embedded Class II.
At optical and near-infrared wavelengths, the SED rises steeply, indicating
substantial dust obscuration in the line of sight.
In Section 4.5, we will determine the extinction to the immediate vicinity
of the star to be A$_V$ = 12 mag.
Figure~3 also shows that the observed outburst amplitude is strongly wavelength dependent,
being much larger at optical wavelengths than in the infrared.

To determine the luminsity of Gaia19bey during the outburst, we
integrated over the available F$\nu$ values and selected the
flux at the outburst maximum for the Gaia, 2MASS, and WISE wavelengths,
while using the available archival data for longer FIR wavelengths,
even though these measurement were not taken during an outburst.
Assuming the minimal possible distance of 1.4 kpc, this method
gives a lower limit to the luminosity during outburst of
$\approx$182 {$L_\odot$}, placing Gaia19bey
in the luminosity range of typical Herbig Ae/Be stars
\citep{Hillenbrand.1992.ApJ.397.613H}. Considering the
uncertainty in the distance, an order of magnitude higher
luminosity is within the range of possibilities.

\subsection{Color-Color Diagram Locus and Variation}

\begin{figure}[h]
\begin{center}
\includegraphics[angle=0.,scale=0.65]{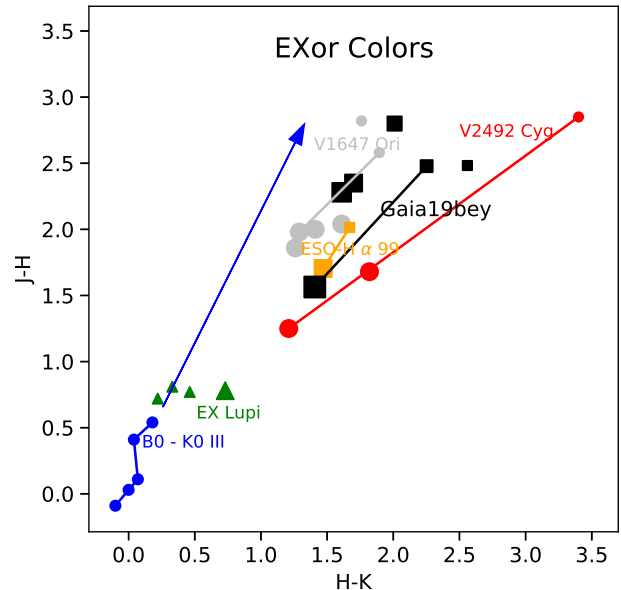}
\caption{
Color-Color Diagram of Gaia19bey at six different epochs, compared
to other EXor-like outburst objects. 
We use the photometry from 2MASS, UKIDSS, the recent WFCAM photometry, and the photometry derived from the
flux calibrated spectra presented here.
We indicate the brightness qualitatively by
the size of the marker, with the 2MASS data point being the largest. 
The black line connects the brightest data point (2MASS) and the most recent
UKIRT/WFCAM measurement from 2020 June 20.
The errors in these measurements are smaller than the size of the symbols
used.
The small blue filled circles represent the locus of unreddenend class III stars
of spectral types B0III to K0III
from \citet{Wegner.2014.AcA.64.261W}
and the blue arrow represents the reddening vector due to interstellar
extinction based on the data by
\citet{Straizys.2008.BaltA.17.125}.
Gaia19bey, in the six different states where data are available,
shares the same locus in this color-color diagram as ESO-H$\alpha$~99
\citep{Hodapp.2019.AJ.158.241.ESOHa99} and a few other deeply embedded
YSO outbursts such as V1647Ori
from \citet[and references therein]{Aspin.2011.AJ.142.135A}
and V2492 Cyg
\citep{Lorenzetti.2012.ApJ.749.188}.
We have
included the prototypical EX Lupi
data points 
from \citet{Herbig.2008.AJ.135.637H}.
in quiescence from the photometry by
\citet{Glass.1974.MNRAS.167.237.EXLupi.phot} and
\citet{Hughes.1994.AJ.108.1071.EXLupi.phot} and the
2MASS catalog, all taken in quiescence, and the
photometry during outburst by
\citet{Juhasz.2012.ApJ.744.118.EXLupi.phot}
to illustrate the point that Gaia19bey and similar object
are indeed much redder than the prototypical EX Lupi.
All the deeply embedded EXors are bluer in the bright state.
}
\end{center}
\end{figure}

\citet{Herbig.2008.AJ.135.637H} has found that
eruptive young stars of the EXor type occupy a characteristic
locus in the $J-H/H-K$ color diagram (their Fig. 10), with redder colors than
what main-sequence stars obscured by interstellar extinction would have.
\citet{Lorenzetti.2012.ApJ.749.188}
has shown a similar color-color diagram including all known
EXors at the time.
While classical EXors, including the prototypical EX Lupi,
are only moderately reddened, many of the recently discovered 
deeply embedded stars with EXor-like spectra
are much redder and occupy the upper right area in Figure~8. 
All these deeply embedded EXors for which $J$, $H$, and $K$ photometry in different
phases of their outburst is available change their position
in the color-color diagram in the sense of being bluer when
bright. 
The variation path in the color-color diagram is flatter than
the interstellar extinction vector. If the variations were to 
be interpreted as solely the effect of extinction variations,
this would indicate a less wavelength dependent extinction law,
consistent with larger grains than are prevalent in the interstellar
medium.
However, we will show in Sections 4.4 and 4.5 that the outburst of Gaia19bey
is not solely caused by extinction variations.
The newly discovered Gaia19bey shares the color-color diagram
locus and its variation with 
other deeply embedded outburst stars studied by
\citet{Lorenzetti.2012.ApJ.749.188} and is closely
similar to 
ESO-H$\alpha$~99, studied by
\citet{Hodapp.2019.AJ.158.241.ESOHa99}
While these ``new EXors'' 
\citep{Lorenzetti.2012.ApJ.749.188} share the basic characteristic
of having an emission line spectrum with classical EXors, their
outbursts last longer, they are of higher luminosity, and their
locus in the color-color diagram is markedly redder.

\subsection{Emission Lines}

In the optical light curve in Figure~2 we have indicated
the times when the spectra were taken with the same color
coding as used in Figures~5 and 6 for the spectra.
The first infrared spectrum (shown in red in Figure~5), taken on 2019 June 29 (MJD 58664), was
taken just one month after the maximum brightness of Gaia19bey during
its present outburst 
at a $G$ magnitude of 18.0.
The second spectrum (shown in green in Figure~6) on 2019 October 23 (MJD 58779)
shows an intermediate stage of the brightness decline 
($G$ = 18.5)
after the maximum when the brightness was rising again to a minor
temporary spike. The emission lines were fainter at that time than in the
first spectrum
both in absolute terms and relative to the continuum.
The change in line strength (Table 3) was more pronounced for the 
\ion{Na}{1} and \ion{Mg}{1} doublet lines than for the hydrogen lines.
The last spectrum shown in blue in Figure 6, taken with SPEX on 2020 March 6 (MJD 58915) when the Gaia magnitude
of 20.0 mag
was about 3 magnitudes below the maximum at optical wavelengths, has changed fundamentally,
showing an almost pure continuum spectrum with emission lines being an order of magnitude fainter
than on 2019 October 23.
The most pronounced emission lines in
this spectrum are the forbidden shock-excited lines presumably emerging in
the outflow at some distance from the star, and apparently not directly
affected by the variability.

We have measured
the equivalent width of several emission lines (Table 3) and plot them against the continuum
flux at 2190 nm in Figure~9. This particular wavelength was chosen to be free of detectable emission
lines and therefore representative of the continuum component. 
In this plot, extinction variation due to dust absorption
would not change the equivalent width of emission lines shining through that extinction,
leading to horizontal lines in this plot. In contrast to this, the equivalent widths
strongly decline with declining continuum brightness of the object, indicating that
the strong photometric outburst from MJD 57500 to 59000 was caused by a substantial
increase in magnetospheric accretion, leading to an EXor-type emission line spectrum.

\begin{deluxetable*}{lcrrrrrr}
\tabletypesize{\scriptsize}
\tablecaption{Emission Lines Strengths}
\tablewidth{0pt}
\tablehead{
\colhead{Line} & \colhead{Wavelength} & \colhead{20190629 Flux} & \colhead{20190629 EW} & \colhead{20191023 Flux} & \colhead{20191023 EW} & \colhead{20200306 Flux} & \colhead{20200306 EW} \\
\vspace{-0.6cm} \\
& \colhead{   nm} & \colhead{10$^{-18}$ Wm$^{-2}$} & \colhead{pm} & \colhead{10$^{-18}$ Wm$^{-2}$} & \colhead{pm} & \colhead{10$^{-18}$ Wm$^{-2}$} & \colhead{pm}
}
\startdata
Pa$\beta$   & 1282 & 6.4 $\pm$ 1.3 & -234 $\pm$ 49 & 3.5 $\pm$ 0.8 & -246 $\pm$ 57 & 0.9 $\pm$ 0.5  & -165 $\pm$ 94 \\
\ion{Mg}{1} & 1504 & 18.3 $\pm$ 0.8 & -341 $\pm$ 15 & 4.5 $\pm$ 0.6 & -158 $\pm$ 21 & 0.5 $\pm$ 0.2 & ~-41 $\pm$ 17 \\
Br 11       & 1681 & 6.9 $\pm$ 1.0 & -100 $\pm$ 15 & 2.7 $\pm$ 0.8 & ~-65 $\pm$ 18 & 0.4 $\pm$ 0.1 & ~-22 $\pm$ ~7 \\
\ion{Mg}{1} & 1711 & 4.9 $\pm$ 0.4 & ~-68 $\pm$ ~6 & 0.9 $\pm$ 0.3 & ~-21 $\pm$ ~8 & 0.2 $\pm$ 0.1 & ~-11 $\pm$ ~5 \\
Br 10       & 1736 & 10.0 $\pm$ 0.9 & -133 $\pm$ 12 & 4.0 $\pm$ 0.6 & ~-89 $\pm$ 13 & 0.2 $\pm$ 0.2 & ~~-7 $\pm$ ~6 \\
H$_2$ S(1)  & 2122 & 2.9 $\pm$ 0.3 & ~-25 $\pm$ ~3 & 2.7 $\pm$ 0.2 & ~-38 $\pm$ ~3 & 5.2 $\pm$ 0.2 & ~-117 $\pm$ ~5 \\
Br 7        & 2166 & 16.6 $\pm$ 2.2 & -143 $\pm$ 19 & 4.4 $\pm$ 1.3 & ~-60 $\pm$ 18 & 0.9 $\pm$ 0.2  & ~-19 $\pm$ ~4 \\
\ion{Na}{1} & 2207 & 11.5 $\pm$ 0.5 & ~-97 $\pm$ ~5 & 3.0 $\pm$ 0.3 & ~-39 $\pm$ ~4 & 0.5 $\pm$ 0.3 & ~~-9 $\pm$ ~5 \\
CO 2-1      & 2294 & 94.4 $\pm$ 0.6 & -770 $\pm$ ~5 & 30.8 $\pm$ 0.4 & -377 $\pm$ ~5 & 2.2 $\pm$ 0.3 & ~-40 $\pm$ ~6 \\
\enddata
\end{deluxetable*}


The optical spectrum taken with SNIFS at the UH 88" telescope (Figure~5)
was taken on 2019 December 9 (MJD 58826) at around a $G$ magnitude of 19, between the last two infrared
spectra both in time and in object brightness.
It clearly shows the forbidden line of [\ion{O}{1}] at 630.0~nm and weakly indicated the 636.3~nm~ line
The 
H$\alpha$ emission is strong and the 
blended [\ion{S}{2}] 671.6/673.1~nm are weakly indicated.
In combination these lines are
characteristic of Herbig-Haro shocks. 
In addition, the \ion{Ca}{2} triplet is very strong in emission
as is often found in T Tauri stars as discussed, e.g., by \citet{Muzerolle.1998.AJ.116.455M}.
The ratio of these 3 lines
is expected to be 
1:9:5 proportional to their (gf) values in the optically thin
case, as discussed by 
\citet{Azevedo.2006.AA.456.225.CaIIinTTauri} for the case of
classical T Tauri stars.
In Gaia19bey, the line ratio is much closer to unity, indicating
that the conditions in the line emitting region are close to
optically thick. This is similar to the pattern of line
strength found by 
\citet{Hamann.1992.ApJS.82.247.TTauri}
and \citet{Hamann.1992.ApJS.82.285.CaIIHerbigAeBe}
for T Tauri and Herbig AeBe stars in that the \ion{Ca}{2} line
at 849.8 nm has the highest line flux.
The only absorption line detected in the optical spectrum is
the \ion{Li}{1} line at 670.8~nm (the insert in Figure~5). 
The lithium line is an indicator of young age in stars since in convective
atmospheres, it is being transported sufficiently deep into the stellar
interior to be destroyed by nuclear reactions, as described in the review by
\citet{Pinsonneault.1997.ARAA.35.557.mixinginstars} and the references therein.

\begin{figure}[h]
\begin{center}
\includegraphics[angle=0.,scale=0.53]{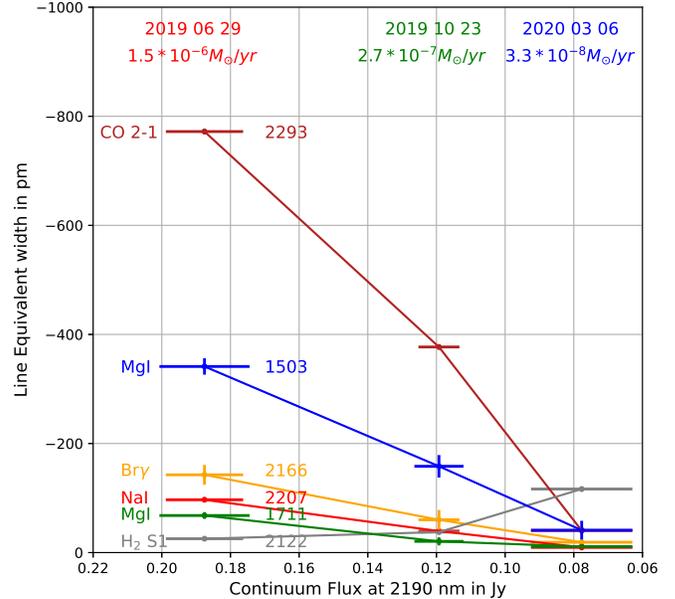}
\caption{
Plot of emission line equivalent width against the continuum flux at a wavelength of 2190 nm.
The x-axis is in order of decreasing continuum flux, to roughly match the historic evolution of
the observed part of the light curve. The dates of the 3 epochs of infrared spectroscopic observations
are indicated in the top, in the same color coding as in Figures 2 and 6.
Varying foreground continuum extinction does not change the equivalent width of spectral lines
and in this scenario, the equivalent width plots of all spectral lines would be horizontal lines.
Constant emission line flux on top of declining continuum will lead to a positive slope in this
diagram, as is the case for the H$_2$ S(1) line emission.
For all the other emission lines, this diagram indicates that the line equivalent width
strongly declines with decreasing continuum flux, i.e., that the emission line strength
declines disproportionally stronger than the continuum.
}
\end{center}
\end{figure}

\subsection{Accretion Luminosity}
We have used two infrared emission lines of hydrogen to measure
the line luminosity and deduce an estimate for the extinction
and the accretion luminosity.
For Herbig AeBe stars, less obscured analogs to Gaia19bey in
terms of luminosity, \citet{Fairlamb.2017.MNRAS.464.4721.Lacc_vs_Lline}
have recently measured both UV continuum emission,
a direct indicator of accretion onto the stellar surface, and
the line luminosities of many optical and near-infrared emission
lines, and have given empirical relationship between these
line luminosities and the total accretion luminosity.
Gaia19bey is in the luminosity range of Herbig AeBe stars,
and we used these L$_{acc}$ vs. L$_{line}$ relationships to
derive L$_{acc}$ for two strong hydrogen emission 
lines (Br$\gamma$ and Br 11-4) in our Keck OSIRIS
spectra from 2019 June 29 and 2019 October 23. 
As a side note,
\citet{Fairlamb.2017.MNRAS.464.4721.Lacc_vs_Lline}
do not give the calibration for Br 10-4, so that we
could not use this strong line.
Used naively, their calibration 
procedure ignores the effect of extinction. We can get an
estimate of the extinction, and correct the accretion luminosity
for extinction by assuming that the two emission lines, when
properly corrected for extinction, should give the same accretion luminosity.
We used the continuum extinction law measured by
\citet{2011.ApJ.729.92.cloud.extinction}
in lines of sight through dense molecular
cores as the best available estimate of the extinction through
the edge-on protostellar disk of Gaia19bey.
For the 2019 June 29 data when the emission lines were strong, we solved for the extinction, 
expressed as A$_K$ when the accretion luminosity determined
by each line becomes the same and obtained an A$_K$ = 1.59 $\pm$0.6
and $log(L_{acc})$ = 1.39 $\pm$0.35.
For the 2019 October 23 data, when the emission lines were substantially
weaker and the determination of the extinction much more difficult,
we assumed the same extinction as on 2019 June 29: A$_K$ = 1.59,
and took the average of the accretion luminosities measured in Br$\gamma$ and Br 11-4
as the best estimate listed in Table 4.

\begin{deluxetable}{llcl}
\tabletypesize{\scriptsize}
\tablecaption{Accretion Rates}
\tablewidth{0pt}
\tablehead{
\colhead{Date} & \colhead{L$_{acc}$} & \colhead{dM/dt} & \colhead{A$_K$}\\
\vspace{-0.6cm} \\
& \colhead{L$_{\odot}$} & \colhead{M$_{\odot}$yr$^{-1}$} & \colhead{mag}
}
\startdata
\vspace{-0.2cm} \\
2019 06 29 &  $25.^{+31}_{-11}$ &  $1.5^{+1.9}_{-0.7}\times10^{-6}$ &   1.59 $\pm$ 0.6\\
\vspace{-0.2cm} \\
2019 10 23 &  ~$5.8^{+2.8}_{-1.8}$ &  $3.5^{+1.7}_{-1.1}\times10^{-7}$ &   1.59 assumed\\
\vspace{-0.2cm} \\
2020 03 06 &  ~$0.65^{+0.22}_{-0.15}$ &  $3.9^{+1.3}_{-0.9}\times10^{-8}$ &   1.59 assumed \\
\vspace{-0.2cm} \\
\enddata
\end{deluxetable}

The uncertainties in these measurements are large, partly
due to the faintness of the lines, the less-than-perfect subtraction
of the hydrogen absorption lines in the spectrum of the telluric
absorption standard star, and due to uncertainty over the extinction
law. Despite all this, we conclude that shortly after the maximum of the light
curve, Gaia19bey has a similar accretion luminosity and mass accretion
rate than the higher end of the distribution of mass accretion rates in Herbig AeBe stars, 
whose accretion has been studied by
\citet{Mendigutia.2011.AA.535.99.HAeBe.accretion}.
The accretion rate is much lower than that of FU Orionis outbursts
where dM/dt = $1.\times10^{-4}$ M$_{\odot}$yr$^{-1}$ is possible
\citep{Audard.2014.prpl.conf.387A}.
For an approximate conversion of the A$_K$ extinction into the
commonly used A$_V$, 
we use A$_V$ = 7.5 * A$_K$, corresponding to R$_V$ = 5
\citep{Cardelli.1985.AJ.90.1494.opticalextinction}.
Our extinction measurement of A$_K$ = 1.59 corresponds to A$_V$ = 12 $\pm$ 5.

\subsection{Spatial Distribution of Shock-Excited Emission}

The adaptive optics OSIRIS data cubes were used to extract
continuum subtracted images of the emission in the [\ion{Fe}{2}]
line at 1644~nm, and the H$_2$ 1--0 S(1) line at 2122 nm in Figure~7.
For the S(1) line, we have two images in the 50 mas and 100 mas
per spaxel spatial scales. The [\ion{Fe}{2}] emission is extended
over $\approx$ 1 \arcsec and shows bifurcating band of low emission
across the position of the unresolved continuum source. This morphology
strongly suggests that we look at a region of extended [\ion{Fe}{2}]
emission in an edge-on disk system with the disk plane oriented at P.A. $\approx$ 125\arcdeg. 
Similar morphologies were found to be common in young outflow sources when
imaged at high spatial resolution with HST by \citet{Padgett.1999.AJ.117.1490.EdgeOnDisks}.
Morphological features related
to outflow activity are also seen in most other deeply embedded EXors and FUors.
The S(1) line shows very different
features. At the 50 mas scale, the S(1) emission is only slightly
extended to the NW, in the same direction as the dark lane in [\ion{Fe}{2}].
The wider view in the 100 mas scale, the right panel in Figure~7,
shows what looks like the inner walls of an outflow cavity.
The western side of this outflow cavity is not included in our 
field of view, but in combination with the absorption lane
seen in the [\ion{Fe}{2}] image (left panel in Figure~7)
it is consistent with the axis of the outflow cavity being
oriented at P.A. 215\arcdeg.

We have recorded both the [\ion{Fe}{2}] lines at 1256 and 1644 nm, from
which, in principle, a value for the extinction can be calculated.
However, the lines are very faint in our spectra, 
and for the
OSIRIS spectra, the close proximity of the [\ion{Fe}{2}] 1644 nm with the H~Br~12 emission
line, and consequently the excess noise associated with the correction of the telluric
standard absorption line, is problematic. 
We have therefore decided to forego this
exercise. In any case, as Figure~7 shows, the [\ion{Fe}{2}] 1644 nm
emission arises in an extended region around the star 
and a dark lane
bisecting this extended emission indicates absorption in an edge-on disk,
as discussed in Section 4.5.

The presence of an edge-on disk is also consistent with the flat-spectrum characteristics
of its SED, which indicates the presence of a substantial amount
of relatively cold dust around the object.  In young stars, such cold dust
is almost inevitably in the form of a disk.

We have imaged the wider field around Gaia19bey with the UKIRT WFCAM 
in search of more shock-excited outflow features and
the continuum-subtracted image (Figure~4) does indeed show a H$_2$ jet with
several knots and shock fronts. While this jet lies in the general
direction of the opening of the outflow cavity seen in the adaptive
optics image (Figure~7), it does not align well with Gaia19bey,
and, more importantly, the morphology of the shock fronts suggest
that this jet is moving in the general SW to NE direction, i.e., 
in the general direction towards Gaia19bey. We cannot determine the source of this jet with
certainty, but a plausible candidate source is the WISE object
J204008.40+464708.0, 6.0 arcmin SW of the jet, that is detectable out to the WISE band 4,
and is associated with faint S(1) line emission closer to it.
This object is not included in the SIMBAD data base and no records
exist in the literature.
The important point is that Gaia19bey is not associated with
a large scale H$_2$ S(1) jet, outside of the features seen
in the adaptive optics images very close to the star.
As was mentioned in Section 2, there is also no H$\alpha$ emission
that could be morphologically associated with Gaia19bey.

\subsection{Comparison with Similar Objects}

Both the optical spectrum (Figure~5) and the first two of the infrared spectra
in Figure~6 show the characteristic emission lines
seen in very young EXors. 
The hydrogen emission lines are generally assumed to originate
in the accretion funnels in the magnetosphere
of the accreting star, where material flows onto the star along
magnetic field lines. 
The CO bandhead emission is also produced in the hot innermost
parts of the accretion disk by irradiation from the accretion
hotspots near the surface of the star
\citep{Calvet.1991.ApJ.380.617.CODiskEmission}.
In the infrared, the emission
lines are on a strong continuum indicating substantial
emission from hot dust probably irradiated and heated
by the accretion hotspots.

While not two eruptive YSOs have identical characteristics,
it is useful to put Gaia19bey in the context of similar objects.
Gaia19bey is a fairly high luminosity, deeply embedded object in a distant
star-forming region. 
An interesting comparison case is V346 Norma, for which
\citet{Kospal.2020.ApJ.889.148.V346Nor}
have recently published a detailed spectrum and have discussed 
changes of the spectrum over the past several decades.
V346 Nor has undergone a long-lasting outburst from 1980 to 2010,
a relatively brief minimum, and for most of the past decade, a rise
back close to the previous outburst level. While the timescales
of its variations are an order of magnitude longer than for
Gaia19bey, the dip in brightness before the rise to the present
maximum appears qualitatively similar.

A spectrum of V346 Norma obtained in 1983 by
\citet{Reipurth.1985.AA.143.435.V346Nor.spectrum}
showed only an absorption spectrum.
Less than one year later, in 1984
Graham and Frogel 1985
obtained a spectrum showing
weak Li absorption, but also a strong H$\alpha$ emission
line with a P Cygni profile. 
Later in the outburst of V346 Nor, in 1993, an infrared spectrum by
\citet{Reipurth.1997.AA.327.1164.V346Nor}
detected emission of Br$\gamma$, NaI (2206 nm)
FeI (2240nm) and CO bandhead emission, similar to what we have
observed for Gaia19bey during its outburst.
However, the latest infrared spectrum from 2015
\citep{Kospal.2020.ApJ.889.148.V346Nor}
do not detect any of these lines, and the CO bandhead may
be absent or weakly in absorption.
Their spectrum is dominated by a multitude of [FeII] and H$_2$ emission
lines originating in shocks, possibly indicating the emergence of
a new HH object or jet. Otherwise, the spectrum is essentially a
continuum, similar to what we observe in Gaia19bey in the declining
phase of its present outburst.

Another object with characteristics similar to Gaia19bey is
the eruptive variable PV Cep.
The infrared spectrum obtained by \citet{CarattioGaratti.2013.AA.554.66.PVCep}
shows many emission lines, both permitted lines, in particular of atomic
hydrogen, similar to those seen in Gaia19bey and probably originating in
magnetospheric accretion. In addition, many forbidden lines of [FeII]
and H$_2$ were detected and indicate outflow activity, similarly, but
with stronger forbidden lines than in Gaia19bey.
However, \citet{Kun.2011.MNRAS.413.2689.PVCepExtinctionVariations}
have demonstrated that the brightness variations in PV Cep cannot
solely be explained by accretion variations, but that variations
in the extinction also play an important role.

The eruptive young object V1647 Ori, sometimes referred to as McNeil's Nebula
(reference) and being the prototype of the ``MNOR'' class of deeply embedded
high luminosity EXor that
\citet{ContrerasPena.2017.MNRAS.465.3039C} have defined, shared many characteristics
with Gaia19bey:

1) have SEDs of class I or 
are flat-spectrum sources, 
2) show outburst durations $>$1.5 yr but shorter than those usually
associated with FUors, 
3) show spectroscopic characteristics
of eruptive variables, i.e. CO emission or absorption, 
and
4) usually have 2122 nm H$_2$ emission.
Also, their sample contains several objects of high bolometric
luminosity in the range of hundreds to thousands of
{$L_\odot$}, the same range as Gaia19bey, considering the
substantial uncertainty of its distance.
5) Similar to at least one object in their sample, Gaia19bey
changed from an emission spectrum to a continuum during 
the declining phase of the outburst.

\citet{Aspin.2009.ApJ.692.67.V1647Ori.reemerge}
derived a mass accretion rate of $4.6\times10^{-6}M_{\odot}yr^{-1}$
during the outburst of V1647 Ori from the Br$\gamma$ line, and 
\citet{Aspin.2008.AJ.135.423.V1647Ori.quiet} obtained
$1.0\times 10^{-6}M_{\odot}yr^{-1}$
during quiescence. 
They also point out the substantial role that extinction
variations in the scattering path of the reflection nebula must
play in explaining the light curve of this object.
While we do not detect a reflection nebula near Gaia19bey in continuum, 
the prevalence of extinction variations in other YSOs strongly suggests
that such variations play a role in explaining brightness variations
in phases of the light curve when magnetospheric accretion, evident
by emission lines, has already subsided.

The spectral changes observed in Gaia19bey are similar to those
observed by \citet{Guo.2020.MNRAS.492.294.VVV.spectra.var} in at least two
YSOs of the MNOR type in their sample. Their object VVVv374 changed from an emission line
$K$-band spectrum to absorption of Br$\gamma$ over the course of roughly
one year, associated with a $\approx$ 0.5 mag increase in brightness during
a longer brightening phase. Their VVVv662 went
from Br$\gamma$ emission to a pure continuum spectrum after a 
1 magnitude drop in $K$-band magnitude.

Changes in the spectrum of Class I protostars were studied by
\citet{Connelley.2014.AJ.147.125.SpecVar}
in a sample of non-eruptive objects. Even without major
outburst events, changes
in emission line ratios were routinely found.
However, only one object in their sample,
IRAS 03301+3111 showed a strong increase in the
near-infrared continuum and a transition from CO in absorption
to emission over the course of 3 years.

\section{Summary and Conclusions}

We have reported observations of the recent
outburst of Gaia19bey.
The outburst duration of Gaia19bey of $\approx$ 4 yr
is longer than typical EXor outbursts, but much less
than FUor outbursts and places it between
those traditional classes, but it is certainly
closer to EXors than to FUor.
Gaia19bey shares the NIR colors and
mid-to-far IR SED with the most deeply embedded
objects in the broadly defined EXor class.
Outbursts of Gaia19bey appear to be repetitive:
The 2MASS catalog photometry, in comparison with
UKIDSS photometry, strongly suggests a prior outburst
of similar amplitude sometime around 1998.

The optical spectrum and the two infrared spectra taken in the bright phase 
of the present outburst show many emission lines, in
particular \ion{H}{1},\ion{Ca}{2}, \ion{Mg}{1}, \ion{Na}{1} line and CO bandhead emission, 
making Gaia19bey also spectroscopically similar to those 
deeply embedded EXor outbursts.
The third spectrum, taken well past the maximum in the declining phase of the outburst,
shows a substantial change in the spectral characteristics. 
All emission lines
are an order of magnitude fainter,
except the forbidden shock-excited
lines that the adaptive optics images show are emitted in nebulosity
surrounding the star.
The color-color diagram indicates redder colors as the outburst 
was fading and can be interpreted as a cooling of the continuum
component of the spectrum.
In the shock-excited H$_2$ 1--0 S(1) emission line, Gaia19bey shows the morphology of an
outflow cavity, strongly suggesting that it is an active outflow source.
The [\ion{Fe}{2}] line emission at 1644 nm is extended and shows indications of a disk
seen nearly edge-on, with an orientation consistent with the outflow cavity structure
seen in the S(1) line.

At the time of writing this paper, the outburst of Gaia19bey appears to be nearing
its end, but the pre-outburst brightness has not yet been reached again. The
future evolution of the spectrum should be monitored and the quiescent state
of Gaia19bey needs to be studied. In particular, it remains to be seen whether
a photospheric absorption spectrum will emerge in the quiescent phase,
when the dust has sufficiently cooled.

\acknowledgments
The OSIRIS data presented herein were obtained at the W. M. Keck Observatory,
which is operated as a scientific partnership among the California Institute of Technology (Caltech),
the University of California and NASA.
The Observatory was made possible by the generous financial support of the W. M. Keck Foundation.\\
The SPEX data in this paper were obtained at the Infrared Telescope Facility, which is operated by
the University of Hawaii under contract NNH14CK55B with the National Aeronautics and Space Administration.\\
The Pan-STARRS1 Surveys (PS1) and the PS1 public science archive 
have been made possible through contributions by the Institute for Astronomy, 
the University of Hawaii, the Pan-STARRS Project Office and their partner institutions, 
the Max-Planck Society and its participating institutes, 
the Max Planck Institute for Astronomy, Heidelberg and the Max Planck Institute for Extraterrestrial Physics, Garching,
The Johns Hopkins University, Durham University, the University of Edinburgh, 
the Queen's University Belfast, the Harvard-Smithsonian Center for Astrophysics, 
the Las Cumbres Observatory Global Telescope Network Incorporated, 
the National Central University of Taiwan, the Space Telescope Science Institute, 
NASA under Grant No. NNX08AR22G, and
issued through 
the Planetary Science Division of the NASA Science Mission Directorate, 
the NSF Grant No. AST-1238877, 
the University of Maryland, 
Eotvos Lorand University (ELTE), the Los Alamos National Laboratory, and the Gordon and Betty Moore Foundation.\\
This work has made use of data from the ESA mission
{\it Gaia}
\footnote[5]{\url{https://www.cosmos.esa.int/gaia}}
and processed by the {\it Gaia}
Data Processing and Analysis Consortium (DPAC,
\footnote[6]{\url{https://www.cosmos.esa.int/web/gaia/dpac/consortium}}
and the Photometric Science Alerts Team.
\footnote[7] {\url{http://gsaweb.ast.cam.ac.uk/alerts}}
Funding for the DPAC
has been provided by national institutions, in particular the institutions
participating in the {\it Gaia} Multilateral Agreement.\\
ATLAS observations and this work were supported by NASA grant NN12AR55G.\\
This publication makes use of data products from the Near-Earth Object Wide-field Infrared Survey Explorer (NEOWISE), 
which is a joint project of the Jet Propulsion Laboratory/California Institute of Technology and the University of Arizona. 
NEOWISE is funded by the National Aeronautics and Space Administration.\\
Based on observations obtained with the Samuel Oschin 48-inch Telescope at the Palomar Observatory 
as part of the Zwicky Transient Facility project. 
ZTF is supported by the National Science Foundation under Grant No. AST-1440341 
and a collaboration including Caltech, IPAC, the Weizmann Institute for Science, 
the Oskar Klein Center at Stockholm University, the University of Maryland, 
the University of Washington, Deutsches Elektronen-Synchrotron and Humboldt University, 
Los Alamos National Laboratories, the TANGO Consortium of Taiwan, the University of Wisconsin at Milwaukee, 
and Lawrence Berkeley National Laboratories. Operations are conducted by COO, IPAC, and UW.
This work uses archival data obtained 
by AKARI, a JAXA project with the participation of ESA.
This work made use of the ADS, Simbad, and VieziR.\\
This work is based in part on data obtained as part of the UKIRT Infrared Deep Sky Survey and 
additional
near-infrared imaging data from the WFCAM at the UKIRT observatory operated by the University of Hawaii.
This publication makes use of data products from the Two Micron All Sky Survey, which is
a joint project of the University of Massachusetts and IPAC/Caltech, funded by NASA and NSF.\\
This paper makes use of data obtained as part of the INT Photometric H$\alpha$ Survey of the Northern Galactic Plane 
(IPHAS, www.iphas.org) carried out at the Isaac Newton Telescope (INT). 
The INT is operated on the island of La Palma by the Isaac Newton Group 
in the Spanish Observatorio del Roque de los Muchachos of the Instituto de Astrofisica de Canarias. 
All IPHAS data are processed by the Cambridge Astronomical Survey Unit, 
at the Institute of Astronomy in Cambridge. 
The bandmerged DR2 catalogue was assembled at the Centre for Astrophysics Research, 
University of Hertfordshire, supported by STFC grant ST/J001333/1.\\
M.A.T. acknowledges support from the DOE CSGF through grant DE-SC0019323.\\
We thank the referee for constructive comments that helped improve the paper.

\vspace{5mm}
\facilities{ATLAS, Gaia, WISE, Keck:I, IRTF, PS1, UH:2.2m, UKIRT}



\end{document}